\documentclass[prl,twocolumn,showpacs,preprintnumbers,amsmath,amssymb]{revtex4}
\usepackage{graphicx}
\usepackage{dcolumn}
\usepackage{bm}

\usepackage{graphics,epsf,psfig}
\usepackage{amssymb,amsmath,amsfonts}
\usepackage{bbm}

\begin{document}
\title{Quantum phase transition in easy-axis antiferromagnetic
integer-spin chains.}
\author{Massimo Capone$^{1,2}$, Sergio Caprara$^2$, Luca Cataldi$^2$}  
\affiliation{$^1$ Enrico Fermi Center, Roma, Italy\\
$^2$ Dipartimento di Fisica - Universit\`a di Roma ``La Sapienza'',\\
and  Istituto Nazionale per la Fisica della Materia (INFM) - SMC and UdR Roma 
1, \\
Piazzale Aldo Moro 2, I-00185 Roma, Italy}

\begin{abstract}
Antiferromagnetic Heisenberg integer-spin chains are characterized by a 
spin-liquid ground state with no long-range order, due to the relevance of 
quantum fluctuations. Spin anisotropy, however, freezes quantum fluctuations,
and the system is magnetized in the presence of a sufficiently large 
easy-axis anisotropy. We numerically investigate the case $S=1$, by means of 
the density-matrix renormalization group, and find that the freezing of the 
spin liquid into a N\'eel spin solid, with increasing easy-axis anisotropy, is 
a continuous quantum phase transition. Numerical evidence indicates that the 
transition is not in the two-dimensional Ising universality class.
\end{abstract}
\pacs{75.10.Jm, 75.30.Cr, 75.40.Mg}
\date{\today}
\maketitle


Quantum fluctuations play a crucial role in antiferromagnetic (AFM) Heisenberg 
spin chains, as they are strong enough to suppress magnetic long-range order 
at zero temperature. Despite the common absence of a spontaneous staggered 
magnetization in the ground state, the physical properties are, however, 
different for integer and half-integer spin $S$ \cite{hal}. Indeed, 
half-integer-spin chains are characterized by a gapless excitation spectrum, 
power-law spin-spin correlations, and a divergent linear response to a 
staggered magnetic field, and are therefore usually considered as nearly 
ordered. The spectrum of integer-spin chains is instead gapped in the 
thermodynamic limit, with a finite linear response to staggered magnetic 
fields. The ground state is a spin liquid with finite spin-spin correlation 
length, and the gap to the first excited state is usually called Haldane gap.

In this paper we present a numerical investigation of integer-spin (namely
$S=1$) chains, for which the presence of different energy scales makes a 
unified theoretical approach problematic \cite{varietheo}. This difficulty is 
mirrored by a richer variety of physical regimes with respect to 
half-integer-spin chains. Physical effects which freeze quantum fluctuations 
are expected to destroy the spin-liquid phase, possibly leading to a 
magnetization. In a previous paper \cite{noi} two of us focused on the 
freezing of the spin liquid induced by a staggered magnetic field, and found 
that the nature of the spin-liquid ground state entails a peculiar 
non-monotonic dependence of the staggered magnetization on the system size in 
the region of small fields (i.e., whose characteristic energy scale is smaller 
than the Haldane gap), and that the full freezing of quantum fluctuations into 
a polarized N\'eel ``crystal'' is only achieved at fields larger than the 
exchange coupling constant, with a large crossover region at intermediate 
fields.

Here, as another effect suppressing quantum fluctuations, we consider an 
anisotropy in the AFM coupling. Since the Haldane gap is small with respect to 
the exchange coupling, a small (but finite) anisotropy is expected to freeze 
the spin liquid. Thus a zero-temperature phase transition to an ordered phase 
should take place upon increasing the anisotropy. We will focus 
on easy-axis anisotropy: When the anisotropy is large, the
system approaches the Ising limit, magnetized along the easy axis.
Since the two limiting fixed 
points, which correspond to the disordered and fully ordered phases, are 
characterized by different dynamical exponents $z$, the issue arises of 
determining the evolution from the $z=1$ (relativistically invariant)
description of the spin liquid within the non-linear $\sigma$ model in the
isotropic case, toward the $z=0$ Ising limit. 

We base our analysis on the infinite-size density-matrix 
renormalization-group (DMRG) approach \cite{white}, which allows for a very 
accurate description of the ground-state properties of one-dimensional systems.
We provide a complete description of the freezing of the spin-liquid with 
increasing anisotropy, measuring both non-universal 
features, such as the spontaneous staggered magnetization  
and the location of the transition point, and 
universal properties, such as the values of the critical exponents and hence 
the universality class of the transition. The analysis of an easy-plane 
anisotropy, which excludes the appearance of an order parameter, leaving the 
way open to a Kosterliz-Thouless phase transition, is technically and 
numerically more involved, and is currently under investigation 
\cite{noifuture}.

Our starting Hamiltonian is a simple modification of the Heisenberg (exchange) 
Hamiltonian to describe a system with easy-axis anisotropy,
\begin{eqnarray}
\label{hamiltonian}
{\cal H} &=& J\sum_{r=1}^{L-1} \left[ S_r^z S_{r+1}^z + p\left(
S_r^x S_{r+1}^x + S_r^y S_{r+1}^y\right)\right]\nonumber\\
&-&H \sum_{r=1}^L (-1)^{r}S_r^z,
\end{eqnarray}
where ${\bf S}_r=(S_r^x,S_r^y,S_r^z)$ is the spin  operator on site $r$, 
$S$ is integer, and $L$ is the number of sites in the chain. $J>0$ is 
the AFM Heisenberg coupling, $0 \le p \le 1$ is the easy-axis anisotropy 
parameter \cite{notanis}, and $H$ is the amplitude of the staggered magnetic 
field along the $z$ (easy) axis, which is coupled with the staggered 
magnetization. As it is customary and convenient in DMRG 
calculations, we assume open boundary conditions. The two limiting 
cases $p=1$ and $p=0$ correspond to the Haldane spin liquid and Ising 
antiferromagnet respectively. It is worth noting that the model 
(\ref{hamiltonian}) is invariant for $p\rightarrow -p$, so that our results 
equally apply to a system with a ferromagnetic coupling on the $xy$ plane.
 
As discussed in \cite{noi}, we find it more convenient to
perform a transformation on the Hamiltonian (\ref{hamiltonian}) which 
allows us to make use of the standard implementation of DMRG  
\cite{white}. After a local rotation of the reference frame, 
${\tilde S}_r^z=(-1)^r S_r^z$, ${\tilde S}_r^x=(-1)^r S_r^x$, 
${\tilde S}_r^y=S_r^y$, the Hamiltonian (\ref{hamiltonian}) is re-cast in the 
``ferromagnetic'' form
\begin{eqnarray}
\label{gauged}
{\tilde {\cal H}} &=& -J\sum_{r=1}^{L-1}
\left[ {\tilde S}_r^z {\tilde S}_{r+1}^z +
\frac{p}{2}\left( {\tilde S}_r^{+}{\tilde S}_{r+1}^{+}
 + {\tilde S}_r^{-}{\tilde S}_{r+1}^{-} \right)\right]\nonumber \\
&-& H \sum_{r=1}^L {\tilde S}_r^z,
\end{eqnarray}
where ${\tilde S}_r^{\pm}={\tilde S}_r^{x}\pm {\rm i}{\tilde S}_r^{y}$ are
the rising and lowering operators, and the external field $H$ is now coupled 
to the uniform magnetization ${\tilde S}^z = \sum_r {\tilde S}_r^z$. The 
easy-axis anisotropy corresponds to  $|p|<1$, whereas the case $|p|>1$ 
corresponds to an easy-plane anisotropy. In the following we refer to the 
properties of the transformed model described by the Hamiltonian 
(\ref{gauged}), and in particular to the uniform magnetization along the easy 
axis, $\langle {\tilde S}^z\rangle$. The correspondence with the original AFM 
model (\ref{hamiltonian}) is straightforward.

Differently from the case discussed in Ref. \cite{noi}, where the 
magnetization was forced by the external field, a spontaneous magnetization 
$m_0$ is expected to appear now, as quantum fluctuations are frozen with 
increasing easy-axis anisotropy. Indeed, the limiting results, $m_0=1$ for 
$p=0$ (Ising limit), and $m_0=0$ for $p=1$ (isotropic limit), suggest that 
$m_0$ should decrease with increasing $p$. 
However, contrary to the case of half-integer spin, where the system is 
critical at $p=1$, and is therefore spontaneously magnetized for
arbitrarily small anisotropy, here we expect that the system persists in the 
magnetized phase only up to a critical value $p=p_c<1$, and the 
spin-liquid phase extends 
over a finite region $p_c\le p \le 1$. Since, as we discuss below, the 
transition is continuous (of the second order), the issue arises whether the 
proper dimension of the effective model at the phase transition equals or not
the effective dimension of the relativistically invariant spin-liquid phase 
($d_{eff}\equiv d+z=2$, where $d=1$ is the spatial dimension and $z=1$ is the 
dynamical exponent). It is worth noting that the two 
phases (magnetized and spin liquid) are both characterized by a gapped 
excitation spectrum, although of different origin. The gap in the easy-axis 
phase is related to the discrete symmetry, whereas the Haldane gap in the 
spin-liquid phase is of a more subtle nature.

To characterize the ordered phase and the transition, we
compute the magnetization 
$M(J,p,H;L,N_k)=\langle [{\tilde S}^z ]^{\cal T}\rangle$, where  
$\langle\cdot\rangle$ denotes the expectation value over the DMRG approximate 
ground-state wave function, for the system of linear size $L$, with $N_k$ 
states retained in the density-matrix truncation procedure, and the 
superscript ${\cal T}$ indicates the truncated operators defined in the 
approximate basis. We consider the magnetization per site in units of $S$, 
$m=M/(SL)$ (we always take $S=1$), such that $0\le m\le 1$ for $H>0$. 
$m$ depends on $J$ and $H$ only through $h\equiv H/J$. 
Starting from $m(p,h;L,N_k)$, we perform three 
subsequent extrapolations \cite{juo} to extract the spontaneous magnetization 
and the linear susceptibility:
\par\noindent
(i) We extrapolate the finite-size results to the thermodynamic limit 
$L\to\infty$ for any given number of states $N_k$ and field $h$ by fitting the 
large-$L$ behavior of the magnetization as 
$m(p,h;L,N_k)=m(p,h;N_k)+{\cal A}/L$, where the long-living corrections are 
$O(1/L)$ due to the open boundary conditions. According to 
our previous results \cite{noi}, the linear fit in $1/L$ at small field is 
accurate only if the size $L$ of the chain is larger than roughly twice the 
characteristic length scale $\xi$ of the spin-liquid state ($\xi=6.03$ 
in units of the lattice spacing, for $S=1$ \cite{esseuno}). 
\par\noindent
(ii) Once $m(p,h;N_k)$ is obtained, we carefully check the convergence with 
respect to the Hilbert space, by taking the extrapolation to the limit 
$N_k\to\infty$ in the region where $m$ displays a clear asymptotic dependence 
on $1/N_k$ \cite{juo}, $m(p,h;N_k)=m(p,h)+{\cal B}/N_k$. Up to $N_k=100$ 
states were needed for the smallest field $h=0.0001$, to obtain the asymptotic 
behavior. We emphasize that this procedure represents a direct measure of the 
convergence to the limit of infinite Hilbert space of the expectation values 
of the different observables, and is not equivalent to fix a small truncation 
error for the density matrix.
\par\noindent
(iii) We finally obtain the spontaneous magnetization $m_0$ and 
the linear susceptibility $\chi(p)$ by means of the 
extrapolation for small magnetic fields $m(p,h)=m_0(p)+\chi(p) h$. 

All the above extrapolations are carefully checked for convergence, e.g., with
respect to various extrapolation schemes, and the least-square linear fits
used thereby are chi-square tested. The numerical values that we report below 
are affected by an error on the last digit. The error bars for the data points 
reported in the various figures are always smaller than the symbol size.

The change in the ground-state properties with decreasing $p$ is already 
witnessed by the behavior of $m(p,h;L,N_k)$ at fixed $N_k$ and $h$ as a 
function of the system size $L$ (not shown). 
Due to the presence of a finite characteristic length, a non-monotonic behavior 
characterizes the spin-liquid phase at $p=1$, with a bump for 
$L \simeq \xi$ \cite{noi}. 
Reducing $p$ gradually leads to a 
monotonic behavior as a function of $L$, with much larger bulk 
($L \to \infty$) magnetization. The study of the spontaneous magnetization,
reported below, confirms that the change of behavior is associated with the 
change in the ground state from spin liquid to spontaneously magnetized.
\begin{figure}
\begin{center}
\includegraphics[width=6cm]{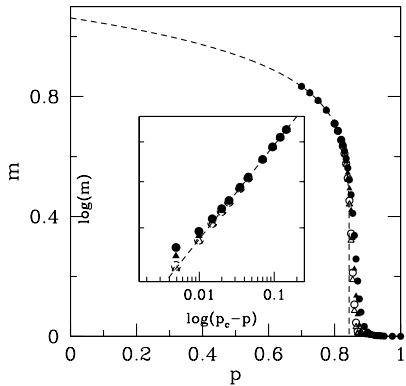}
\caption{The spontaneous magnetization $m$ as a function of the anisotropy
$p$. The different data points refer to different values of the lowest
magnetic field $h_{min}$ used in the small-field extrapolation: 
$h_{min}=0.001$ (black dots), $0.0005$ (full triangles), $0.0001$ (open 
circles), and $0.00005$ (open triangles). The inset displays the same quantity 
in log-log plot to emphasize the power-law behavior. The dashed line is the
fitting curve $m_0=a\left(p_c-p\right)^\beta$ (see text).}
\label{fig1}
\end{center}
\end{figure}

The magnetization $m(p,h)$ increases with decreasing $p$ at each fixed value
of $h$, and naturally increases with the magnetic field $h$ at each $p$. The 
dependence on $h$ is weak at $p\simeq 1$ 
and vanishes as $p\to 0$.
The critical behavior is evident in Fig. \ref{fig1}, where we report $m_0(p)$ 
as a function of $p$, obtained through a zero-field extrapolation
\cite{notahmin}. 
Except for a narrow region around the critical point \cite{notamoltovicino}, 
the data are well described by $m_0=a\left(p_c-p\right)^\beta$, for $p<p_c$, 
and $m_0=0$ for $p>p_c$, with $a=1.09$, $\beta=0.139$, and 
$p_c=0.845$, as obtained by numerically minimizing, with respect to $p_c$, the 
least-square fit of $\log m_0$, which yields an ``optimum'' value of $a$ and 
$\beta$ for any given $p_c$. The inset of Fig. \ref{fig1} shows         
that 11 extrapolated points fall on the fitting curve. 

\begin{figure}
\begin{center}
\includegraphics[width=6cm]{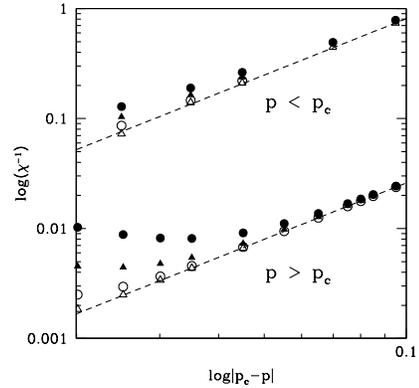}
\caption{Inverse susceptibility $\chi^{-1}$ as a function of $p$           
in a log-log scale. Notations are the same as in Fig. 1. The dashed lines
are the fitting curves $\chi^{-1}=b(p_c-p)^\gamma$ for $p<p_c$, and 
$\chi^{-1}=b'(p-p_c)^{\gamma'}$ for $p>p_c$ (see text).}
\label{fig2}
\end{center}
\end{figure}

We can also extrapolate the linear susceptibility $\chi(p)$ for $h\to 0$ 
(see Fig. \ref{fig2}), finding the power-law behavior 
$\chi^{-1}=b(p_c-p)^\gamma$ for $p<p_c$, and $\chi^{-1}=b'(p-p_c)^{\gamma'}$ 
for $p>p_c$, once again except in a small region close to the critical point.
Using for $p_c$ the value obtained from the spontaneous magnetization, 
$p_c=0.845$, and performing a least-square fit of $\log \chi^{-1}$,
we find $b=41.1$, $\gamma=1.71$, $b'=1.30$, and $\gamma'=1.70$. The numerical 
agreement of the two indices $\gamma$ and $\gamma'$ within numerical accuracy 
is a check of the reliability of our results. When studied over the entire 
parameter range, as a function of $p$, $\chi (p)$ obviously vanishes in the 
Ising limit ($p\to 0$), where the magnetization saturates ($m_0\to 1$), and 
has a finite value in the Haldane spin-liquid phase at $p=1$, which increases 
with increasing $S$ ($\chi=18.50$, for $S=1$ \cite{esseuno}).

\begin{figure}
\begin{center}
\includegraphics[width=6cm]{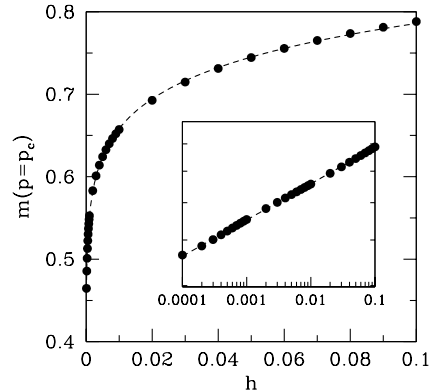}
\caption{The magnetization at the critical point $p=p_c$ as a function 
of the magnetic field. The inset shows the same quantity in log-log scale
to emphasize the critical behavior. The dashed line is the fitting curve 
$m(h,p=p_c)=ch^{1/\delta}$ (see text).}
\label{fig3}
\end{center}
\end{figure}

The exponent $\delta$ which characterizes the behavior of the magnetization
as a function of the external field $h$ at criticality, 
$m(h,p=p_c)=ch^{1/\delta}$, can be calculated by means of the scaling law 
$\delta=1+\gamma/\beta$. Substituting the values of $\gamma$ and $\beta$ 
found above we obtain $\delta=13.4$. We can compare this value with a 
direct measure of  $\delta$. Fig. \ref{fig3} shows the 
data for $m(p=p_c,h)$, with $p_c=0.845$, as a function of $h$. The log-log 
plot shows a linear behavior over three decades with a least-square exponent 
$\delta=13.0$ in agreement with the one obtained from the scaling law within 
numerical accuracy, and $c=0.938$.

\begin{figure}
\begin{center}
\includegraphics[width=5cm,angle=-90]{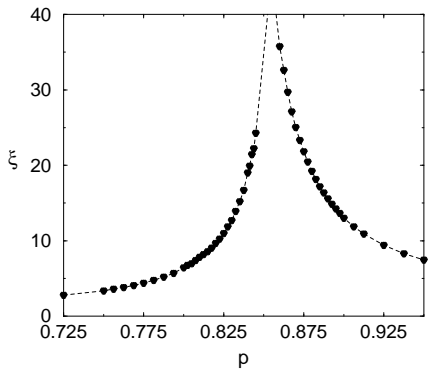}
\caption{Correlation length $\xi$ as a function of the anisotropy $p$. The 
dashed line is a guide to the eye.}
\label{fig4}
\end{center}
\end{figure}

Finally, by calculating the spin-spin correlation function along the easy axis
$C_K(r)=\langle {\tilde S}_{K+3+r}^z[{\tilde S}_{K+3+2r}^z ]^{\cal T}\rangle$ 
($r=0,1,2,...$) \cite{notacorr}, we determine the exponents $\nu,\nu'$ which 
characterize the divergence of the correlation length $\xi$, given by   
$\xi = g(p_c-p)^{-\nu}$ for $p<p_c$, and 
$\xi=g'(p-p_c)^{-\nu'}$ for $p>p_c$. The correlation length is obtained by 
fitting  the long-distance behavior of the correlation function as 
$C_K(r)=A{\rm e}^{-r/\xi}+B$, where  $B=C_K(r\to\infty)$. $\xi$ (see Fig. 
\ref{fig4}). By fitting the data for $\log\xi$ as a function of $\log|p-p_c|$, 
assuming $p_c=0.845$ as determined from the fit of the spontaneous 
magnetization, we find $\nu=0.86$, $g=0.45$ (fitted with 11 points) and 
$\nu'=0.87$, $g'=1.04$ (fitted with 12 points). It is worth noticing that the 
correlation length $\xi$ as a function of the anisotropy parameter $p$ is 
rapidly reduced to a quantity of the order of the lattice spacing as $p$ is 
reduced on the Ising (ordered) side, and reaches the finite value  
$\xi\simeq 6$ in the Haldane spin-liquid state, at $p=1$.

Using the scaling laws, we can determine the effective dimension at the 
phase transition $d_{eff}=(2\beta+\gamma)/\nu=2.3$. We have therefore 
numerical evidence that the fixed-point effective model is not 
relativistically invariant. In the flow from the spin-liquid to the 
fixed point, along with the various critical exponents, also the dynamical 
exponent $z$ is therefore corrected in a non-trivial way.

In summary, we have studied the role of easy-axis anisotropy in the $S=1$
AFM spin chain by means of DMRG.
Our calculations show that the freezing of the quantum 
fluctuations due to increasing easy-axis anisotropy at small magnetic field is 
much alike the one obtained with increasing field \cite{noi}. In particular 
the non-monotonic dependence of the magnetization $m$ on the size of the 
system $L$ observed for the isotropic chain is gradually transformed into 
a monotonic increase. 
The extrapolation to zero magnetic field reveals a quantum phase transition
from a spin liquid to a spin solid at a finite value of the anisotropy
parameter, $p_c=0.845$. The values of the exponents indicate that the phase 
transition falls within a different universality class than the $d=2$ 
Ising universality class (although quite close to it). Thus in the evolution 
from the spin liquid (where $d_{eff}=2$) to the Ising antiferromagnet (where 
$d_{eff}=1$), a fixed point is reached with a different effective 
dimensionality $d_{eff}=2.3$ (i.e., a dynamical exponent $z=1.3$). This result 
indicates that the lattice-renormalization effects and/or the less relevant 
coupling constants neglected in Ref. \cite{schulz}, play a role, even if 
quantitatively small, so that the critical exponents are close to, 
but different from, those of the two-dimensional Ising model.
  
We acknowledge a stimulating conversation with A. Parola, and many fruitful 
discussions with C. Castellani. We also acknowledge financial support of 
INFM/G through PA-G0-4 and of MIUR Cofin 2001, prot. 2001023848.


\begin{thebibliography}{99}
\bibitem{notanis} $p$ measures the ratio between the $xy$ coupling 
($J_{xy}=pJ$) and the coupling along the $z$ (easy) axis ($J_z=J$), and 
decreases with increasing easy-axis anisotropy. A direct measure of the 
anisotropy is provided, e.g., by $(J_z/J_{xy})-1=(1-p)/p$, which vanishes in 
the isotropic limit ($p\to 1$), and diverges in the Ising limit ($p\to 0$).
\bibitem{notahmin} We report data obtained by varying the  
minimum field $h_{min}$ used in the extrapolation.
The value of $h_{min}$ is important only close to the critical point.
\bibitem{notamoltovicino} The divergence of the response functions and of the
correlation length at the critical point makes it impossible to numerically 
access a small region around it, where very long chains
and small magnetic fields would be required for a good extrapolation, implying
a loss of numerical accuracy and/or an increase of the computing time.
\bibitem{notacorr} The correlation function is calculated, as customary within 
DMRG \cite{white}, by first increasing the chain up to a length $2K+2$, where
$K$ is taken reasonably large to reduce boundary effects, and then starting 
the calculation by fixing a spin on one of the two central sites of the chain, 
while the other goes further and further inside the block nearby, as DMRG 
iterations are continued. Convergence is checked by the overlap of the curves 
$C_K(r)$ at various $K$. We used $K=50,100,150,200$ and
explored the range $0\le r \le 100$ to extract the correlation length.
\end{thebibliography}
\end{document}